\documentclass[manuscript,nonacm]{acmart}

\setcopyright{none}
\usepackage{fontawesome5}

\usepackage{colortbl}

\usepackage[dvipsnames]{xcolor}
\definecolor{mygray}{gray}{0.85}

\usepackage{subfig}
\usepackage{float}

\usepackage{tabularx}

\begin{document}


\title{``Here Be Sharks!'': Enhancing Scientific Communication and Analysis through Authoring Interactivity}

\author{Caroline Berger}
\affiliation{%
  \institution{Aarhus University}
  \streetaddress{Department of Computer Science}
  \city{Aarhus}
  \country{Denmark}
}
\email{caroline.berger@cs.au.dk}

\author{Josh Pollock}
\affiliation{%
  \institution{Massachusetts Institute of Technology}
  \streetaddress{CSAIL}
  \city{Cambridge}
  \country{USA}
}
\email{}

\author{Dylan Wooton}
\affiliation{%
  \institution{Massachusetts Institute of Technology}
  \streetaddress{CSAIL}
  \city{Cambridge}
  \country{USA}
}
\email{}

\author{Arvind Satyanarayan}
\affiliation{%
  \institution{Massachusetts Institute of Technology}
  \streetaddress{CSAIL}
  \city{Cambridge}
  \country{USA}
}
\email{}

\author{Clemens Nylandsted Klokmose}
\email{clemens@cs.au.dk}
\affiliation{%
  \institution{Aarhus University}
  \streetaddress{Department of Computer Science}
  \city{Aarhus}
  \country{Denmark}
}

\renewcommand{\shortauthors}{Berger et al.}

\begin{abstract}
We report on a case study for designing authoring environments for interactive visualizations to enhance scientific work. 
We conducted a workshop and prototype review with a group of marine biologists. 
When it came to visualizing their data, participants identified challenges in conveying their research accurately and completely as well as and analyzing it with ease. 
Based on our findings, authoring environments should consider the context of scientific work aspects such as domain expertise, collaboration culture, and publication traditions.
We propose consideration of non-traditional programming languages and environments for scientific work, discuss ways of facilitating interactive visualizations for scientists, and examine ways of meaningfully integrating AI. 
The critiques, artifacts, and reactions from the scientists along with our analysis and discussion inform how computational tools should be designed for scientific work.  
\end{abstract}





\begin{teaserfigure}
    \centering
    \includegraphics[width=\linewidth]{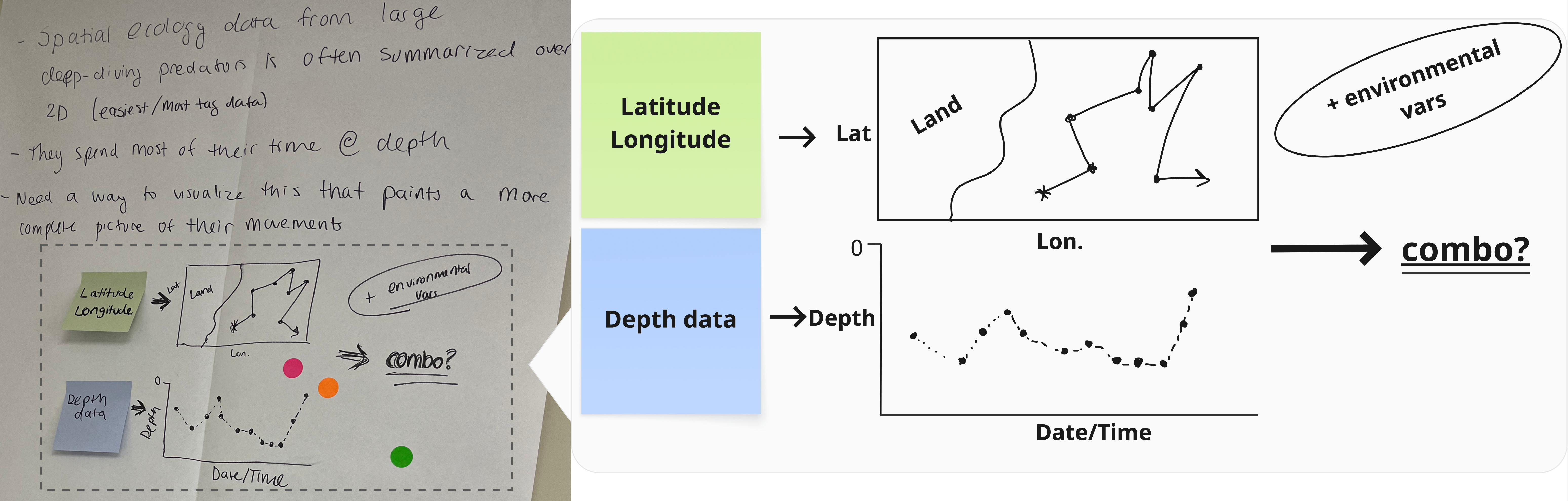}
    \caption{
    \textbf{Telling fisherman where the sharks are.} Post-doctoral marine biologist uses interactive visualizations in videoconferencing presentations with fisherman to answer their questions. Despite wrangling a shark to attach a Fitbit and having the data on hand, she struggles to convey the shark's journey in a visualization. 
    }
    \Description{Figure shows on the left a photo of a large piece of white paper with bullet point hand writing at the top, and a diagram at the bottom. The diagram has a rectangle with dashed lines around it. Next to it, is a digital image. The digital image has two sticky notes one green that has Latitude and Longitude written on it the other sticky note below it that has Depth data written on it. Both sticky notes have arrows pointing to two diagrams. On the top (Latitude and Longitude) there is a Land label and a path with dots and an arrow. The x axis is labeled Lat and the y axis is labeled Long. Below the diagram with Depth data pointing at it, is a plot with Depth on the y axis and Date/Time on the x axis. Inside the plot are points connected by a dashed line. In between the two charts there is an arrow with combo? underlined twice. In the corner is a circle with +environmental vars inside.}
    \label{fig:shark}
\end{teaserfigure}

\maketitle

\section{Introduction}
In the era of computing-infused science, interactivity presents an opportunity to enhance the communication of complex information within the scientific community and to the public at large.
We focus on marine biologists whose wide array of work tasks overlap with the work of scientific professionals from other disciplines. 
For example, the marine biologists that we studied perform lab work, field work, and computer work. 
They dissect fish, dive to retrieve coral samples, and create R scripts to visualize shark movement patterns. 
Their visualization work spans mapping marine species migrations,  analyzing biogeochemical cycles, and depicting starfish habitats.
The marine biologists we worked with report that the traditional static figures they typically make have limited expressive power. 
For the purposes of this work, we define interactivity as the phenomenon when visual elements in software can react to user input like keystrokes or clicks, changing the information presented in a display.
We argue that interactivity offers promising avenues for enhancing scientific communication, exploration, and collaboration.
One participant mentions that the ability to zoom in and out helps her to see finer and broader groupings of species during analysis, a concrete example from our work.

However, creating something as simple as the zoom functionality and authoring interactivity at large remains a significant barrier for scientists.  
Popular environments and frameworks are predominately programmatic such as Vega~\cite{satyanarayan2014declarative}, D3~\cite{bostock2011d3}, and React~\cite{react}. 
Scientists may lack training to leverage these tools~\cite{hannay2009scientists, midas2020nanoscience}.

The overarching research direction that drives this work are the ways scientific work tasks could be changed through interactivity, and the design requirements for tools that support interactive element authorship.
We anchor our study within the oceanography work context, and scope our focus to interactive visualizations with the following specific research questions:\\
\textbf{(RQ1)} How could oceanographic work tasks be meaningfully enhanced through interactive visualizations?\\
\textbf{(RQ2)} What are the design requirements for a tool that supports marine biologists in authoring interactive visualizations?\\
To answer these questions, we met with marine biologists, first in a group as a workshop, then one-on-one as a prototype review. 

Our findings can be translated to other scientific fields that also perform field, lab, and computer work like the marine biologists from our study.
Marine biologists are an ideal group as marine biology doesn't necessarily have a work tradition steeped in teaching advanced computational skills in contrast to fields like computational biology. 

Our thematic analysis revealed that scientists want to convey an accurate and complete picture of their research.
This helps them when communicating with their peers in the scientific community as well as stakeholders who depend on their findings.
In addition to oral communication,  they use visualization during exploratory analysis.
Interactivity has the possibility to reduce chaos and direct attention.
However, we found inaccessible tooling limited the expressive power of scientists.



After reporting on the work tasks that could be supported by interactivity, and considering the design of an authoring environment, we discuss broader issues of computational tools in scientific work practice. 
Our work suggests a departure from traditional code-based editing environments and examines sociotechnical considerations beyond tools, like the chart transformations that are common among certain subdisciplines. 

Our contributions are as follows: (1) we identify the challenges marine biologists face in visualizing their data, (2) we propose opportunities for interactivity to help scientists analyze and communicate their findings, (3) we explore the considerations for an authoring environment for interactive visualizations, and (4) we demonstrate co-design as a method for eliciting visualization challenges.
This paper seeks to uncover the nuances around interactivity authorship in scientific contexts to inform the design of future tools to enhance scientific communication and analysis possibilities.


\section{Related Work}
This work is situated at the intersection of human-computer interaction, scientific programming, and visualization. 
In this related work, we examine the current state of scientific computational work including the practices around visualizing science. 

\subsection{Computer Programming in the Scientific Profession}
Scientists often work in computational notebooks, such as Jupyter and Observable~\cite{perkel2018computational, perkel2021reactive}.
Unlike professional software engineers, scientists prioritize outputs over code quality when programming~\cite{kelly2015scientific, easterbrook2009engineering, berger2024scientists}. 
Practical concerns such as time constraints prevent scientists from performing test-driven development, a widely accepted practice in software engineering~\cite{widder2019barriers}.
Furthermore, Berger and colleagues found that scientists see programming mostly as a tool~\cite{berger2024scientists}. 

In a study with nanoscientists, Nouwens and colleagues cooperatively designed a lab notebook prototype for RNA design~\cite{midas2020nanoscience}. 
The nanoscientists expressed frustration managing their computing infrastructure--the installation, maintenance, and troubleshooting--created roadblocks to their work. 
Working with the tools and scripts without training resulted in a disempowering experience for the scientists. 
This work identified principles of computational media with a scientific user  group--distributability, shareability, malleability, and computability--our work expands on these identified aspects, especially in sociotechnical dimensions. 

Lakier and colleagues found software practices to be marked by fragmentation~\cite{lakier2025understanding}.
The marine science participants often had incomplete understandings of what tools their peers used and why. They developed idiosyncratic and ad hoc approaches to software engineering, resisted adopting unfamiliar tools, and frequently preferred to build solutions from scratch rather than reuse existing code. 
Finally, when tools were developed to be reusable, the tool would rarely travel beyond the single research labs.
The limited scope of available GUI tools often drove scientists toward coding, yet they found programming intimidating, and at times misaligned with the mental models they held of the scientific phenomena under study.
Our work addresses the challenges in the gap between powerful programming-based tools and limiting graphical user interfaces.

\subsection{Practices of Visualizing Science}
Scientists' code often produces visualizations that confirm their expectations, or communicate their findings~\cite{berger2024scientists}. 
Ravi and colleagues' work expands upon this finding~\cite{ravipreliminary}.
In their work with climate scientists, they found that climate scientists used visualizations as a diagnostic tool, and overall, they spend a large amount of their time making and analyzing visualizations.
In an aligned observation with oceanographers, Easterbrook and Johns uncovered that visualization mediates code tasks so that oceanographers can test and validate their work~\cite{easterbrook2009engineering}.
Prior work has also called for a continued investigation into intermediate interactive representations as a way to facilitate collaboration among scientists during their visualization and analysis activities~\cite{berger2024scientists}. 

In interviews with ocean science domain experts, Afzal and colleagues identified interaction needs, such as details on demand and the ability to view various abstraction levels~\cite{afzal2019state}. 
They also identified customizability as a requirement, especially when it comes to preparing presentation of results.
Our work builds upon the interaction needs by adding contextual nuance around interactivity opportunities, and by proposing authoring environment design considerations where customization would occur.

In a longitudinal study with four oceanography teams, Kuksenok and team recounted how a new visualization tool was not adopted~\cite{kuksenok2017deliberate}.
Tableau~\cite{tableau}, a commercially available software, is a graphical user interface that supports the creation of visualizations including interactive elements.
While initially met with excitement, Tableau was quickly abandoned by the lab.
Namely, it failed to help a scientist with a crucial task, creating publication-ready figures due to lack of control over fine-grained formatting details. 
The scientists returned to programming in R~\cite{Rlang}, but adopted some of the visualization choices from Tableau. 
The study reports on the importance of precision and control for scientific users, and also the chasm between text-based traditional programming environments and direct manipulation graphical user interfaces.  
Our work aims build on this work by identifying other needs and characteristics of authoring environments for scientists.  





\section{Method}
With the goal of learning about marine scientists' work, especially as it relates to opportunities for interactive visualization, we deployed a multipart study. 
First, we conducted a workshop in a participatory design style with seven marine biologists. 
Then, we designed a prototype, or \textit{provotype}~\cite{mogensen1992towards}, to be a provocative conversation starter. 
We reviewed the prototype with three of the marine biologists who participated in the workshop to gather additional insights.
The study was conducted in accordance with the ethical approval processes of the involved institutions.
The scenario from the workshop, details about the prototype as well as screenshots of the prototype, and accompanying questions asked during the prototype review are in the supplementary materials.

\begin{figure*}
    \centering
    \includegraphics[width=1\linewidth]{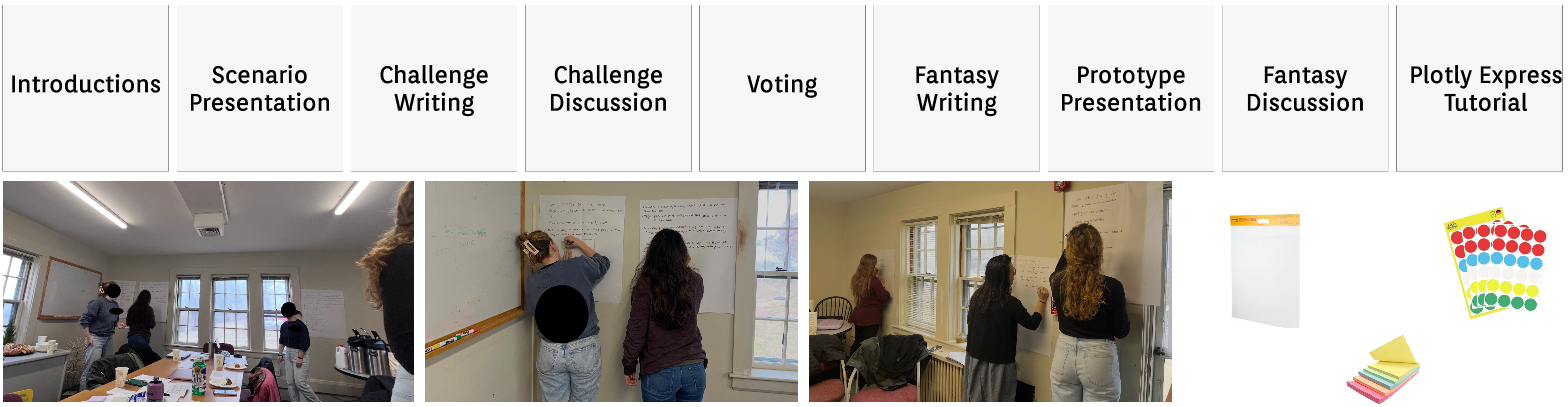}
    \caption{\textbf{Workshop activities.} The workshop included facilitators presenting a scenario and later provocative prototypes. Marine biologists externalized their challenges and fantasy solutions through writing, discussing, and voting.}
    \Description{Figure shows digital sticky notes spanning from left to right: Introductions, Scenario Presentation, Challenge Writing, Voting, Fantasy Writing, Prototype Presentation, Fantasy Discussion, Plotly Express Tutorial. Below the row of sticky notes are three photos of the workshop participants writing on large pieces of paper that are posted on the wall. In the corner are the materials from the workshop: large pieces of paper that can be posted to the wall, sticky notes, and dot stickers.}
    \label{fig:workshop}
\end{figure*}

\subsection{Positionality}
The study was conducted by a team of HCI researchers in large academic institutions based in North America and Scandinavia. 
The lead author is a white North American cisgender woman in her twenties. 
Her prior engagement with ocean science as a high school student influenced her interest in the domain. 
From the research team, two white men in their twenties also facilitated the workshop, and a man based in Scandinavia and a man based in North America provided supervision. 
One co-facilitators was exposed to ocean science at camps as a young person, and the other co-facilitator is an amateur polar scientist, and has experience visualizing and modeling sea-ice data.
As a whole, the team is committed to fostering more inclusive environments in visualization. 
However, we acknowledge there is power and privilege afforded to us based on our identity, role, and educational background.
While we've tried to reduce implicit and explicit biases, to be objective, the positionality of our team as a whole and as individuals influences the decisions in the study design, recruitment, execution of the workshops and prototype reviews, and how we interpret the data.

\subsection{Workshop}
To recruit participants to the workshop, we contacted a marine biologist that is affiliated with our institution. 
After adapting the timing and duration of the workshop based on their advice, we circulated a recruitment blurb. 
Out of fourteen marine biologists from a large oceanographic institution in North America who expressed interest in attending the workshop seven people participated. 
Four PhD students, two post-doctoral affiliates, and one principal investigator participated in an in-person workshop at their place of work.
Table~\ref{tab:participants} shows an overview of the participants.

The workshop followed the future workshop method~\cite{jungk1987future, kensing2020generating}, containing critique, fantasy, and realization phases. 
Our ambition was to collect information about how scientists perform their work and how interactivity could help. 
We used the future workshop method for its mutual learning benefits: we learned about the challenges of scientists, how they envision the ideal future, and introduced them to technological alternatives. 
A scenario served as a frame of reference during critiques.
We reviewed provocative prototypes during the fantasy phase, see figure~\ref{fig:workshop} for overview of activities.
The workshop occurred in early March, 2025 in a conference room at the scientists' workplace.
We gathered informed consent from all participants prior to beginning the workshop.
When describing the workshop in the recruitment and scheduling materials, as well as in the introduction, we framed its focus on interactive visualization. 

\textbf{Scenario} After participants introduced themselves to the others we presented a scenario underscoring a challenging situation about analyzing, presenting, and collaborating on a visualization within a research group (see supplementary material). The scenario was constructed with input from two external subject matter experts in marine science.

\textbf{Critique} We asked participants to use the scenario as a guide.
On large pieces of paper hung up throughout the room scientists wrote down challenging scenarios they face at work.
They shared these with the group, and then each person placed three stickers on which challenges they found the most compelling.

\textbf{Fantasy} They then fantasized on an ideal world by writing on another large piece of paper.

\textbf{Prototypes} To encourage continued fantasy brainstorming, we then presented prototypes to the group, and gathered their reactions.
When choosing prototypes to show the participants, we selected a variety of prototypes that demonstrated some interactivity they may not have seen before.
We reviewed research tools from the visualization and HCI communities and formed a diverse set of authoring environment examples.
Scientists reacted to video clips of Potluck's Plant tracker~\cite{Litt_Schoening_Shen_Sonnentag}, tldraw ``make real''~\cite{Ruiz_2023}, and Ten Brighter Ideas~\cite{Victor_2010} as well as a prototypical lab notebooks for nanoscientists~\cite{midas2020nanoscience}, and semi-formal programming~\cite{ryanSemiFormal}.

\textbf{Tutorial} At the end of the session, we taught scientists about how to create interactive visualizations using plotly express~\cite{plotly}.
We chose plotly express as it supports interaction authoring without extensive programming effort and has wide support for map visualizations, which are common chart types in oceanography~\cite{thomson2014data}.
We designed the tutorial as a learning opportunity for the participants, and didn't collect data during this part.

\subsection{Prototype Review} 
Informed by the reactions to the prototypes during the workshop and comments about computational practices among the participants, such as programming habits, digital toolsets, and data challenges, we developed a prototype (see supplementary material).
We recruited participants from the workshop to review the prototype.
P2, P4, and P7 were available to meet for approximately 45-minutes of in-depth individual sessions.
We gathered informed consent prior to beginning the prototype review.
The review consisted of three parts: an opening semi-structured interview, a demonstration and discussion of the prototype which included the participants interacting with the tool, and a closing semi-structured interview (see supplementary material) . 
The prototype helped us to uncover the marine biologists visualization challenges, details of their context, and their visions for future tools.
The reviews occurred in July 2025.

\subsection{Data Analysis}
The audio and video from the workshop and prototype review were recorded. 
We also took pictures of the artifacts that participants produced during the workshop.
We produced a transcript using OpenAI's whisper~\cite{Openai} and read through the transcripts and then highlighted portions for significance~\cite{braun2021one, braun2006using}. 
We used these significant statements to derive themes, then went back through the transcript to find additional quotes pertaining to the themes.
For example, we defined the theme \textit{Scientists want to communicate the truth} from the workshop transcript excerpt \textit{``I'm potentially missing these ecologically important low abundant taxa''}. 
We also analyzed the participants' drawings and writings, labeling the diagram and bullet points about combining shark position data in figure ~\ref{fig:shark} with the theme \textit{Scientists want to communicate a complete picture}.
The lead author performed the analysis using hard-copy, physical materials including sticky notes, and a whiteboard. 
During the analysis, a supervisory member of the research team provided input on the derived themes, discussing and making any necessary adjustments to the themes.

\begin{table}
    \centering
    \begin{tabular}{lccc}
    \textsc{Participant} & \textsc{Role} & \textsc{Focus} \\
    \midrule
    P1 & PhD Student & Meroplankton\\ 
    \midrule
    P2* & PhD Student& Fisheries \\ 
    \midrule
    P3 & Post-Doc & Marine Snow \\ 
    \midrule
    P4* & PhD Student & Bacteria in Coral Reefs\\ 
    \midrule
    P5 & PI & Microbiology \\ 
    \midrule
    P6 & PhD Student & Bacteria and Ecology\\ 
    \midrule
    P7* & Post-Doc & Marine Predators \\ 
    \end{tabular}
    \caption{\textbf{Participant demographics.} Participant role and field. Astrict (*) denotes participants who also did the prototype review.}
    \label{tab:participants}
\end{table}


\section{Findings}
During the workshop and prototype review, we learned about the types of visualizations that marine biologists make: tree visualizations, stacked bar charts, maps, and network graphs. 
They make visualizations for themselves to perform exploratory analysis, as well as for others in the form of figures for journal articles and slide deck material for meetings with non-scientific stakeholders.
We also learned about the type of information these scientists visualize--communities of species and their relationship to environmental conditions, the abundance of bacteria on corals, and shark movement.
We focused our thematic analysis on: 
(RQ1) How could oceanographic work tasks be meaningfully enhanced through interactive visualizations?
(RQ2) What are the design requirements for a tool that supports marine biologists in authoring interactive visualizations?
As a result, through the workshop and prototype review we identified opportunities for enhancing scientific work through interactivity, and design considerations for authoring environments. 
 
\subsection{Opportunities for Enhancing Scientific Work through Interactivity}
Scientists shared the challenges they face when trying to perform analysis with visualizations and using them to communicate their results to others. 
While some challenges could be overcome by highly technical programming skills and deep visualization expertise, interactivity also provides a promising pathway forward.
Specifically, we identified that interactivity might help scientists convey the scientific meaning of their work. 
In addition to communicating an accurate representation of their scientific findings, interactivity might also help them to share these messages with their broader context. 
Interactivity may also allow for visualizations that extend  beyond immediate relations and in aid scientists in the analysis of their data. 

\subsubsection{Conveying Accurate Science}
By visualizing their data for communicative purposes, scientists aim to convey representative portrayal of their findings.
A coral researcher described a \textit{``rainbow of nothingness''} is produced when plotting her data into a segmented bar chart using R with ggplot as she waits for her old computer to run a program that plots every single taxa. 
As a result, she selects only the top 200 taxa, and consequently:
\begin{quote}
    \textit{``I'm potentially missing these ecologically important low abundant taxa'' (P4, Workshop)}. 
\end{quote}
In the prototype review, she expands on this concern, citing the knock-on effects in her field.
\begin{quote}
    \textit{``one of the limitations of the field is... the inherent assumption  that things that are rare are not ecologically important but that's not necessarily grounded in ecology'' (P4, Prototype Review)}
\end{quote}
The concern was not only about partitioning the data haphazardly, but also how taxonomic resolution might be lost when bacterial species are aggregated. 
As she explained, plotting at higher taxonomic levels could collapse distinctions that were critical to their interpretation:
\begin{quote}
    \textit{``They will plot at a higher taxonomic rate like phylum or class... that would put me and my dog in the same color and we know that me and my dog have different niches [group laughs] so I don't want to lose my fine data at the taxonomy level'' (P4, Workshop)}
\end{quote}
The reflections of the participant illustrate how conveying truth in visualization goes beyond raw data. Proper visualization involves maintaining nuance and avoiding distortions that could give a misleading account of findings, just as
collapsing taxonomic distinctions or discarding rare taxa risks misrepresenting ecological realities. 
However, an overly granular approach including every data point often overwhelms both scientists and their audiences. This sentiment also echoed in the notes of principal investigator P5 and post-doc P3. 

\subsubsection{Depicting a Complete Picture}
Alongside accuracy, participants also expressed the challenge of showing a complete, rich portrayal of their data. 
In many cases, single visualizations risk being too reductionist, leaving out important dimensions of ecological and temporal complexity. 
For instance, one researcher highlighted the difficulty of capturing the movement of predatory fish:
\begin{quote}
\textit{``I think one of our big issues is how [large predatory fish like tuna and sharks] spend so much time at depth that ideally we want a cool way to show their tracks at depth and how they're interacting with oceanographic variables throughout the actual tracks that they're taking'' (P7, Workshop) \& figure~\ref{fig:shark}}    
\end{quote}
The post-doctoral researcher and her team tag sharks by wrestling them on a small boat, to attach a Fitbit to the sharks. 
After wrangling the shark along with the Fitbit data, it is the technical tool kit that limits their expressive power--the ability to tell the story of the shark's path.

Another marine biologist pointed to the multi-layered nature of ecological data—across fish species, seasons, and years—which made it difficult to design visualizations that adequately communicated the breadth of patterns:
\begin{quote}
    \textit{``So, I have all these species, and I have all these seasons, and I have all these years, so how do I show that?'' (P2, Workshop)}
\end{quote}
Like the shark researcher, the fisheries PhD student has the data ready, but no way to present it.

Interactivity has the potential to  combine views of data to communicate the temporal dimensions of the shark's paths, and variations to a fish's habitat throughout the seasons and years.
By allowing researchers to show different dimensions--such as depth, seasonality, or species interactions--interactive systems can surface patterns that static plots obscure facilitating communication with important stakeholders like fisherman and protected area managers.

\subsubsection{Greater Ease during Individual Exploration of Data}
Participants underscored the need for visualizations that facilitate their own exploration of data. When working with large datasets, researchers often faced overwhelming complexity, where visualizations became cluttered rather than clarifying. One participant described this issue vividly:
\begin{quote}
    \textit{``If I plot all of the positions, then it's just thousands of points on a map of 30 different colors... it's chaotic... and I don't have a way of simplifying that yet'' (P1, Workshop})
\end{quote}

When analyzing her data, PhD student P4 notes the difficulty in shifting attention between important parts of the visualization. 
\begin{quote}
    \textit{
    ``The legend is so large  you can't look at the legend and the chart at the same time'' (P4, Workshop) 
    }
\end{quote}

Such experiences illustrate how interactivity could transform the act of data exploration. 
Features like collapsible and interactive legends that highlight data points would allow scientists to focus their attention. 
Rather than choosing between oversimplified graphs or cluttered datasets, researchers could dynamically curate their visual space to focus on the patterns most relevant to their questions.
By keeping scientists ``in the flow'' during analysis, such systems would not only streamline interpretation but also create space for deeper insights to emerge from complex data.

Progressive visualizations~\cite{ulmer2023survey} and dynamic queries~\cite{ahlberg1992dynamic} have been well researched within the visualization community, and pose possible solutions to interactivity challenges faced by the marine biologists with their large datasets.
However, we found that the scientists were not familiar with these techniques.
This underscores the gap between answers that exist in the visualization research community, and the techniques used in marine biology work practice.

\subsection{Design Considerations for Authoring Environments}
Through our work we found that scientific communication and analysis can be improved through interactivity. 
Yet, scientists actually being able to enhance their visualizations through interactivity is a challenge. 
We identified considerations in the design of environments to empower this user group to be able to author interactive visualizations. 

\subsubsection{Expertise of a Scientist User}
The scientist is an expert in their domain, and is often intimately familiar with the data they are trying to visualize.
They often know how the data was collected, often performing the collection themselves, as P4 when she is studying the bacteria on corals. 
\begin{quote}
    \textit{
    ``I'm taking chunks out of corals, and we have a lot of limitations on how we can collect our samples. And so I don't necessarily trust my n of 12 has a robust differential abundance, and I don't feel satisfied with [differential trends].'' (P4, Workshop)
    }
\end{quote}
She has an expert perspective on her data since she gathered it and therefore she has a unique perspective on it's limitations.
This information is crucial for her data analysis, visualization, and sharing.

Years of education and research experience contribute to scientists sense of deep understanding of the domain in which their data is situated. 
Computer tools, especially those with automation and decision support should preserve and promote the scientific expertise of the user.
For example, the system could have mechanisms for the scientists to provide information about the data upfront, and they should have a high autonomy to express information related to their project and discipline. 

At the same time, scientists are not necessarily visualization experts. 
During the workshop, two scientists expressed an explicit need for support:
\begin{quote}
    \textit{``What is the best way to display data? Should I separate it... How do I know which [visualization] to pick?'' (P6, Workshop)}
\end{quote}
\begin{quote}
    \textit{
    ``Sometimes [visualizing data] can be very overwhelming. What do we include? How do I group things?'' (P2, Workshop) 
    }
\end{quote}

Here, authoring environments can play a key role.
Computer systems are well positioned to assist with decision support, and providing the benefits and drawbacks of different visualization choices, and styles, as well as presenting opportunities for interactivity to be added. 
Further, recommendations could consider the shape of the scientists' data, heuristics on visualization best practice, and other user parameters. 

\subsubsection{Destination of the Visualization}
In considering the domain expertise of scientists as it relates to the specifics about their data, project, and discipline, scientific work also has traditions, cultures, and even novel approaches to doing things. All of which can influence design choices of an authoring environment.
As journals and conference articles are the principle medium for many scientists to share their research, this format of static, space limited print documents pose problems for visualization in general, and especially interactivity.

If an author decides to include interactive elements, since they're not tightly coupled and packaged with the official published article, they struggle to get the reader to visit the interactive visualization. 
\begin{quote}
    \textit{``You can include the code [with the paper], but then people would have to go to it.'' (P2, Workshop)}
\end{quote}
Participants imagined ways of embedding interactive visualizations directly into publications, or at least linking them seamlessly:
\begin{quote}
    \textit{``[I want an] easier way to make website with interactive figures that could be included in a paper, would also be cool if at some point journals allow you to include animated and/or interactive figures within the paper itself since most people nowadays read papers online'' (P2, Workshop (written))}
\end{quote}
This tension reflects a broader mismatch: although modern research increasingly generates complex datasets and dynamic figures, dissemination practices remain tied to print-era constraints. As post-doc P7 noted,
\begin{quote}
    \textit{``The level of data that we're capable of producing now and the figures that we can make, they still have these limits as if it's being printed, but it's all online. Some of them will have e-supplementation, some of them will have documents and stuff'' (P7, Workshop). 
}
\end{quote}

Scientists imagined how interactivity could help overcome these limitations by supporting deeper engagement with the data:
\begin{quote}
    \textit{``It would be really cool to have an interactive component where people could, when you're reading a paper, zoom in on a graph, click the points and stuff. I think that would solve a lot of the issues, either with the volume of data that we have, or the nuance of each data point that you lose in the 2D printed version that they're still publishing'' (P7, Workshop)
}
\end{quote}
Others emphasized the potential of alternative formats, such as animations or interactive visualizations, to better convey temporal or high-volume patterns:
\begin{quote}
\textit{``When we're talking about those papers that have a bajillion plots, especially modeling figures, it used to be so much easier if it was a little video. It was showing me through time instead of January, February, March, April, May, June, all plotted, if it was a living plot. And when we do that in presentations, it's like, oh, that makes so much sense. We can't do that.'' (P3, Workshop)
}
\end{quote}
Beyond clarity, participants also noted that interactive and visually striking outputs can generate visibility and excitement, which are tied to professional success. 
As one post-doc explained:
\begin{quote}
    \textit{``Online it would be cool to have something a little bit more interactive. It also generates this buzz. Which then people will read the paper and cite it if you have buzz around it. That's kind of the pressure now. And I think why what you're doing is so important is now when you submit a paper to most journals, but especially the high impact ones like Science, Nature. They always ask you for a graphical abstract. And that can really make or break a paper like that. You get judged on that so quickly if it's on Twitter or Blue Sky or Facebook or Instagram or whatever. And you see it, you're like either immediately yes or immediately no.'' (P7, Prototype Review)}
\end{quote}
Taken together, these perspectives highlight how the destination of visualizations not only constrains what scientists can do, it also creates its own pressures and presents different opportunities. 
Authoring environments must therefore support both the practical realities of current publishing and the aspirational practices scientists hope to see adopted in the future.

\subsubsection{Easy Sharing}
Participants emphasized the difficulty of sharing interactive visualizations with peers, collaborators, and broader audiences. 
While code notebooks, websites, and presentations are common dissemination channels, each comes with barriers that limit accessibility and ease of use. 
As one participant expressed, scientists need 
\begin{quote}
    \textit{``More streamlined ways of creating shared notebooks where people could give you specific feedback (kind of like a google doc)'' (P2, Workshop, (written))
    }
\end{quote}
She pointed to the increasing prevalence of websites for collaborative data sharing:
\begin{quote}
\textit{``Websites are becoming more common to show fisheries data for sharing between colleagues'' (P2, Workshop, (written))}    
\end{quote}
However, participants struggled with integrating interactive outputs into traditional formats such as slide decks:
\begin{quote}
    \textit{``I had a class presentation last spring that I wanted to
show my interactive plot and I gave up because I couldn't figure out how to put it into a powerpoint presentation'' (P2, Prototype Review)}
\end{quote}
The recurring theme was that the lack of straightforward mechanisms for packaging and distributing interactive visualizations hinders their use, even when the technical capacity to build them exists.
\begin{quote}
    \textit{``There has to be a quick way to publish it or to make it into a shareable form.'' (P2, Prototype Review)}
\end{quote}
These reflections suggest that authoring environments should prioritize seamless pathways for sharing visualizations, whether through embedding in presentations, exporting to common formats, or publishing to lightweight web platforms. 
Much like the evolution of word processors and collaborative editing tools, visualization systems could benefit from features that enable feedback, versioning, and broad accessibility without requiring additional programming expertise. 
By lowering the barriers to sharing, interactive visualizations can move beyond individual use and become more powerful instruments for scientific collaboration, education, and communication.

\section{Discussion}
Our study highlights how marine biologists struggle communicating and interacting with their data and how interactivity in visualization potentially can reshape scientific communication and analysis. 
Marine biologists expressed the dual challenge of conducting sensemaking with large, multidimensional datasets while also needing to convey results to peers, stakeholders, and the scientific community. 
Through the workshop and prototype review, we identified opportunities and systemic constraints shaping how interactivity can be adopted in scientific work. 
We discuss broader implications of our work below.

\subsection{Co-Design for Eliciting Visualization Challenges}

In this study, we demonstrate that co-design is a promising method for uncovering visualization challenges that marine biologists face.
Specifically, we employed the future workshop method, scenarios, and prototypes. 
Co-design and participatory design methods at large have been employed in a wide range of human-computer interaction projects, and within visualization specifically, Jänicke and team identified early prototyping, as examples of concrete manifestations of ideas, to be beneficial for generating feedback from ecologists and humanities scholars~\cite{janicke2020participatory}.

In our case, the sequence of the future workshop, of first challenges then fantasy, as well as the setup for participants to write down their responses, and vote on each others, helped us to capture individual, as well as shared challenges.
Employing a concrete scenario served as helpful framing to scope and focus the discussion.
By presenting real life examples of authoring environments for interactivity, we were able to spark continued ideation, the prototypes served as a conversation starter for the marine biologists. 
Using participatory design methods such as the future workshop method, prototypes, and scenarios can increase the richness of collected information, uncover shared experiences, and create opportunities for learning among participants.


\subsection{Non-Traditional Environments for Authoring}

\begin{figure}
    \centering
    \begin{minipage}{0.48\linewidth}
        \centering
        \includegraphics[width=\linewidth, height=4cm, keepaspectratio]{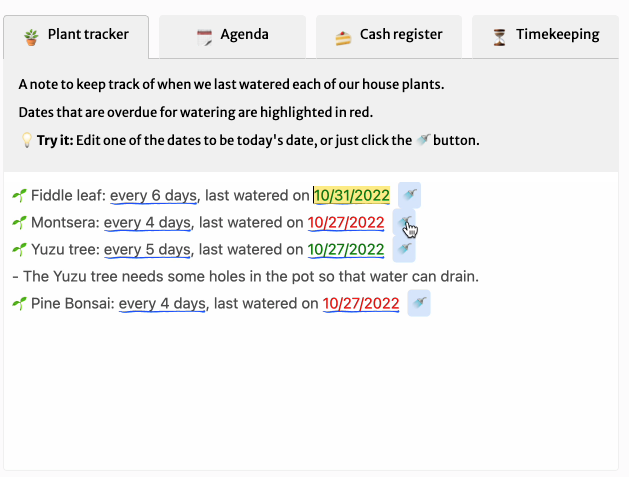}
        \caption{\textbf{Potluck Plant Tracker.} The plant tracker could be adapted for tracking fish feeding demonstrating a potential use case of malleable software in scientific work.}
        \Description{Figure shows several tabs, Plant tracker is highlighted and Agenda, Cash register, and Timekeeping are not highlighted. The Plant tracker tab says ``A note to keep track of when we last watered each of our house plants. Dates that are overdue for watering are highlighted in red. Try it: Edit one of the dates to be today's date, or just click the shower button. There are five rows of a plant icon, name of a plant: every x days, last watered on xx/xx/xxxx and a shower icon. Some dates are green some are red. The mouse is hovering over the shower icon.}
        \label{fig:potluck}
    \end{minipage}
    \hfill
    \begin{minipage}{0.48\linewidth}
        \centering
        \includegraphics[width=\linewidth, height=4cm, keepaspectratio]{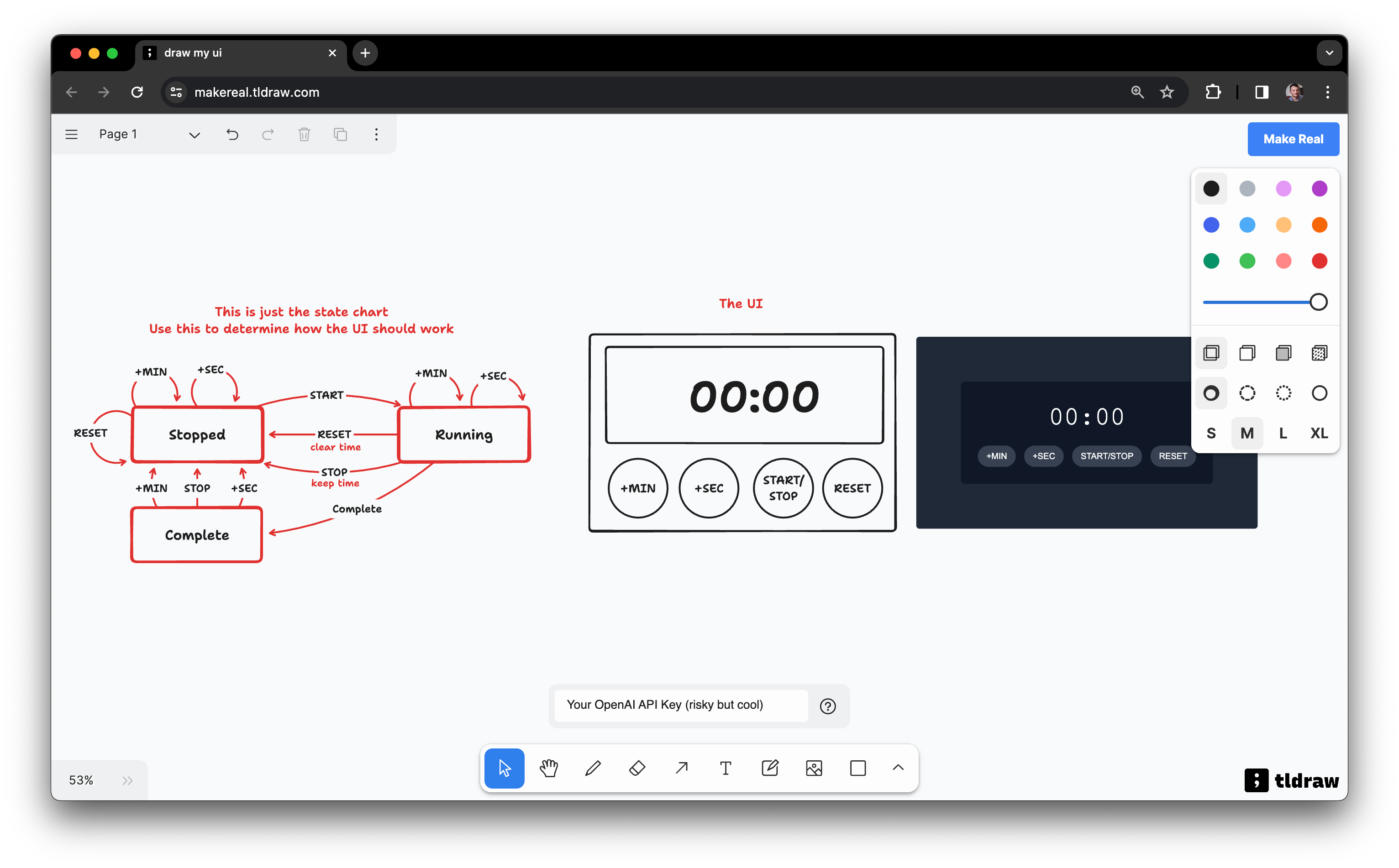}
        \caption{\textbf{tldraw.} A timer created using the tldraw. On the canvas, users can sketch a diagram and interface and instantly generate a working prototype.}
        \Description{Figure shows a web page with a canvas. On the canvas there is a state diagram in red on the left side and on the right a sketch of a timer and next to it a polished digital interface of a timer. At the bottom of the canvas is a tool bar, and on the right hand side a button that says make real as well as color and shape selection tools.}
        \label{fig:tldraw}
    \end{minipage}
\end{figure}

Moving beyond Adobe Illustrator, RStudio, and Jupyter Notebooks, the traditional environments scientists author visualizations, malleable software, canvas editors, and semi-formal environments are examples of non-traditional computing with the potential to empower scientist in analyzing their data and communicating their findings with possibilities for enhancement to their work.
Malleable software aims to empower users to reshape their digital tools, allowing them to adapt technology to their needs rather than adapting themselves to rigid systems.
Instead of swapping between Illustrator, R, and Excel, as described by the scientists during the workshop and prototype review, malleable software makes an amalgamation possible.
Unlike a traditional application-centric model that treats users as passive consumers, malleable software invites users to become creators and editors, enabling custom workflows~\cite{malleablesoftware}.
One example is the \textit{Potluck Plant tracker} prototype~\cite{Litt_Schoening_Shen_Sonnentag} (see figure~\ref{fig:potluck}) which uses freeform text as a source of information for computations, and entangles interactive elements into generated text, using a base programming language of Javascript. 
\textit{Potluck Plant tracker} aims to enable users to turn static digital notes into interactive and useful personal mini tools.
During the workshop, scientists enthusiastically discussed how the \textit{Potluck Plant tracker} prototype could be adapted for their work—for example, coordinating fish feeding schedules. 

This type of customizable and interactive system directly addresses real pain points in their day-to-day research:
\begin{quote}
    \textit{``People who have live fish in the lab... keeping track of that  especially if it's a shared thing too. So you can keep track ... when I [fed fish] during my master's I had a notebook out [in the lab]... [it would be helpful] if you have 10 different fish that all have to eat'' (P4, Workshop)}
\end{quote}
Moving beyond interactivity applied to visualization, malleable software creates the opportunity to personalize tools so that they can assist in scientific work, like shared task management. 
One drawback to the \textit{Potluck Plant tracker} is that it requires Javascript programming proficiency, which is an unlikely skill among the marine biologists, future work should explore mechanisms of tailoring malleable software that doesn't require a high level of programming expertise.

Scientists reacted positively to the visual, direct manipulation interface of the \textit{tldraw}~\cite{Ruiz_2023} tool (see figure~\ref{fig:tldraw}).
\begin{quote}
    \textit{``Making the plot represent your data, not so hard, but making [the plots] look nice, I feel like that's a really nice thing.'' (P1, Workshop)
    }
\end{quote}
While programming in R and Python, scientists edit lines of code to change presentation details.
During this process they need to translate shades of color, distances, and label names into code, run it, evaluate the result, and repeat. 
This requires moving directly to the formal specification of R or Python syntax. 
In contrast, canvas environments invite experimentation through quick feedback to presentation changes.
These environments also contain instruments, like color selector tools, which can help the user to explore alternatives.
For fine-tuning and iterative editing of presentation details, environments that incorporate direct manipulation, design-like tools, and a canvas could lessen translation work~\cite{arawjo2020write} and increase expressive opportunities for science communication. 

Pollock and team examined the representation mismatch that leads to tedious translation work while programming and proposed a design space for building new languages and tools~\cite{pollock2024designing}. 
They recommend considering the dimensions of the environment, for example support for spatial representations like a diagram on a canvas editor or the enforcement of linguistic representation of lines of R code in RStudio.
According to Pollock and colleagues, the aim of semi-formal programming environments is to enhance expression possibilities by matching the representations with the goal and task of the user taking into consideration structure, level of detail, and concreteness of the representations supported by the environment.

The canvas environment of tldraw resonates with what diSessa~\cite{disessa_changing_2001} refers to as \textit{computational media} in his call to abandon the confines of applications for intellectual and creative work on computers. 
A possible future of scientific software is semi-formal programming combined with diSessian computational media that mitigating the fragmentation that~\cite{lakier2025understanding} identified.

\subsection{Expressing Marine Biology in Code}
\begin{figure}
    \centering
    \includegraphics[width=0.6\linewidth]{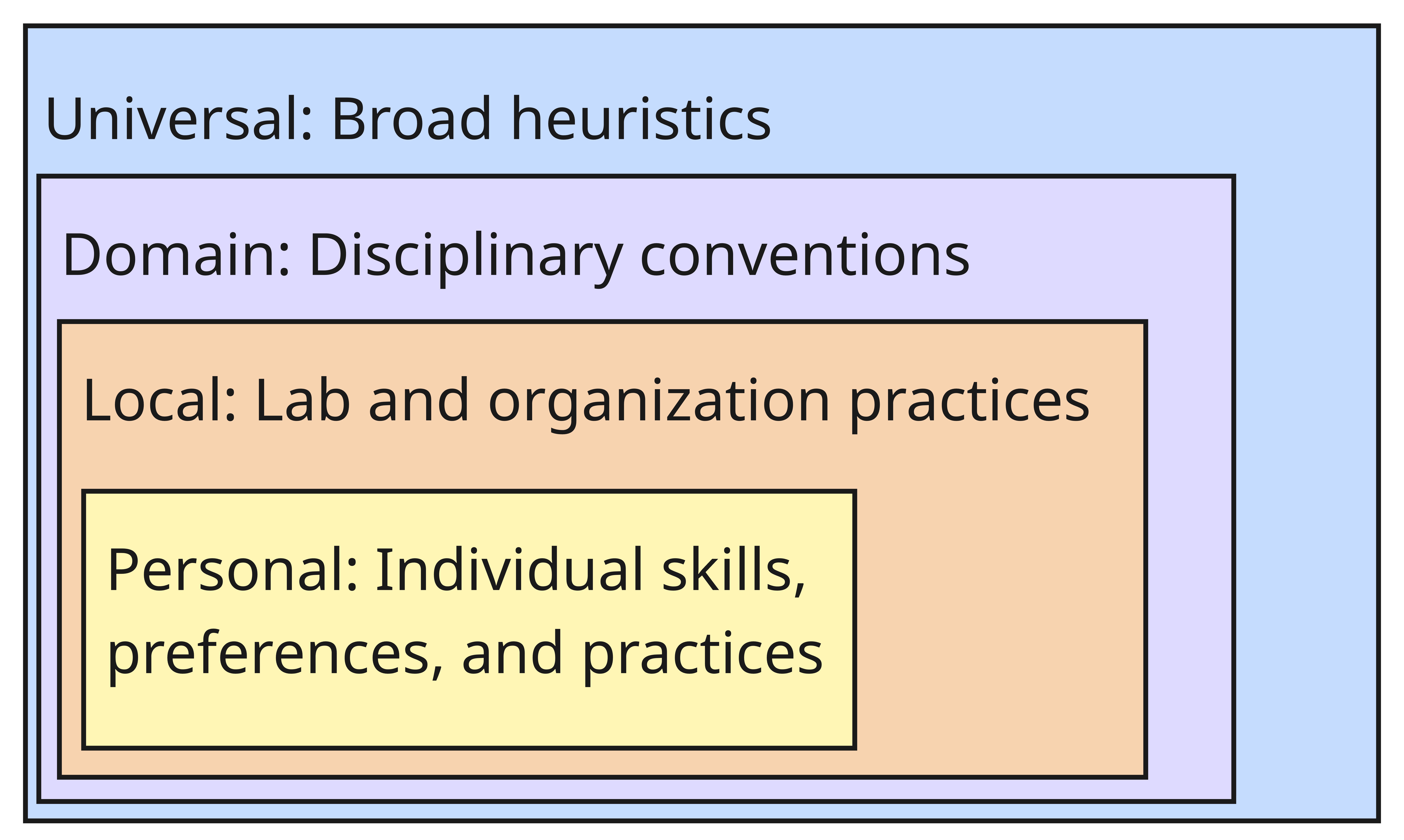}
    \caption{\textbf{Scientific tool-context fit.} When integrating interactive authoring tools into scientific workplaces, universal, domain, local, and personal contextual information is helpful in identifying or building the right authoring environment for interactivity.}
    \Description{Figure shows a diagram with four nesting rectangles in several colors. The outermost to innermost read: Universal: Broad heuristics, Domain: Disciplinary conventions, Local: Lab and organization practices, Personal: Individual skills, preferences, and practices}
    \label{fig:toolfit}
\end{figure}

It important to acknowledge that even if a marine biologist is trained in programming as computer scientists, the gap between reasoning about biological phenomena and expressing those in code is a source of breakdowns, mental overhead, and significant context switches.
In the vein of bridging the gap, we propose an analytical model, inspired by Munzner's Nested Model of Visualization Design and Validation~\cite{munzner2009nested}, to learn more about scientists' context in relation to their tooling.
The Nested Model contains domain situation, data/task abstraction, visual encoding/interaction idiom, and algorithm, different levels where visualization research can occur. 
Our work is aligned with describing the domain situation and strives to help match authoring environments to users. 
The model is a continuum of support that ranges from hyper specific contexts to more general areas. 
We have four levels: universal, domain, local, and personal as shown in figure~\ref{fig:toolfit}. 
Universal relates to heuristics on visualization and data types which can be applied broadly. 
For example, a bar chart is better than pie charts for comparison.
Then, there is the domain.
Disciplinary conventions define the domain, for ocean scientists, there is a practice of flipping the y axis: 
\begin{quote}
    \textit{
    ``We do a lot of the reversed y-axis because we're showing the surface of the ocean at the top'' (P2, Prototype Review)
    }    
\end{quote}
Even more specific, are the local customs.
Each lab has an overarching research area, such as bacteria or arctic environments. 
In this vein they have a scientific community that they publish their work in, and stakeholders that they talk to, like local fisherman, protected area managers, or members of local, state, and federal government.
\begin{quote}
    \textit{
    ``We also work with protected area managers who are used to looking at scientific stuff and would prefer to have somebody give them a three-sentence summary, they don't want to look at graphs''(P7, Prototype Review)
    }
\end{quote}

In terms of the personal level, the languages people use varies, even diverging from their supervisors and close colleagues in the case of the marine biologists we spoke with who said:
\begin{quote}
    \textit{
    ``People in my lab use MATLAB, Python, R, all sorts of different languages.'' (P2, Workshop)
    }
\end{quote}
They also have different workflows, with P7 using illustrator frequently to edit figures produced originally with R in RStudio.
Collecting information about these dimensions can help with evaluating the tool-context fit. 

Assessing the match between the criteria at each level from the scientific user's context and the capabilities of a tool is a promising step towards finding tools that will empower scientists in their computational work.
Tools must align with the layered realities of scientific work--from accepted rules to domain traditions, lab customs, and individual workflows.

\subsection{Limitations}
While our findings can translate across other scientific disciplines some adaptations might be required.
We expect that scientific fields, like computational biology, that have a strong computing tradition will differ in their authoring environment needs. 
Another limitation lies in our choice to use of a provocative prototype rather than a fully functional, deployed tool.
With a tool, scientists could give feedback based on their usage in their work context, reflecting on the tool's mismatch or uncovering additional insights from tool-work frictions.

\subsection{Future Work}
Future work includes developing authoring environments that balance flexibility with guidance, supporting scientists who are experts in their domain but may lack advanced programming or visualization skills.
These tools should be informed by the context of scientists. 
Beyond tool development, examining the broader implications of interactivity in scientific practice is critical, particularly in how these tools change the analysis and communication of science.
We recommend future work include citizen scientists as well as communities organizing around marine system stewardship.

\section{Conclusion}
In this work, we present a case study exploring the design of authoring environments for interactive visualizations to support scientific work, based on a workshop and prototype review with marine biologists. 
Participants highlighted challenges in accurately conveying and effectively analyzing their data, revealing the need for tools that account for the broader context of scientific work. 
Drawing on these insights, we argue for moving beyond traditional programming environments and propose approaches to facilitate interactive visualizations that align with scientists' contexts.
Our work surfaces a gap--- sophisticated tools exist for authoring interactive visualizations, however in their current form, they are inaccessible to scientists.

\begin{acks}
    We thank the anonymous reviewers for their feedback on this paper.
    Any opinions, findings, and conclusions or recommendations expressed here are those of the authors and do not necessarily reflect the views of the funding agency.
    We thank Ryan Yen for his feedback in the early stages of the project, and to Meredith Chou for the input she provided during analysis.
    A big thank you to Benjamin Carrier for his helpful suggestions and Benjamin Gregory for his review of the paper from a biology terminology perspective. 
    Andreas Muenchow and Henry Churchill Henson provided subject matter expertise on oceanography informing the design of the study.  
    We thank the marine biologists for their time, enthusiasm, and for the important science they do to keep our oceans healthy.    
\end{acks}

\bibliographystyle{ACM-Reference-Format}
\bibliography{scientists-programming}

@String{addrACM           = {New York, NY, USA}}

@String{addrIEEE          = {Piscataway, NJ, USA}}

@String{jourTVCG          = {{{IEEE} Transactions on Visualization and Computer Graphics}}}

@String{procCHI           = {Proceedings of the {ACM} Conference on Human Factors in Computing Systems}}

@String{procUIST          = {Proceedings of the {ACM} Symposium on User Interface Software and Technology}}

@String{pubACM            = {{ACM}}}

@String{pubIEEE           = {{IEEE}}}

@incollection{kensing2020generating,
  title={Generating Visions: Future Workshops and Metaphorical},
  author={Kensing, Finn and Madsen, Halskov},
  booktitle={Design at work},
  pages={155--168},
  year={2020},
  publisher={CRC Press},
  doi={https://doi.org/10.1201/9781003063988-10}
}

@book{jungk1987future,
  title={Future Workshops: How to Create Desirable Futures},
  author={Jungk, Robert and M{\"u}llert, Norbert},
  year={1987},
  publisher={Institute for Social Inventions}
}

@inproceedings{lakier2025understanding,
  title={Understanding Marine Scientist Software Tool Use},
  author={Lakier, Matthew and Irwin, Andrew and Vogel, Daniel},
  pages={1--14},
  year={2025},
  doi={https://doi.org/10.1145/3706598.3713621},
  booktitle=procCHI,
  address=addrACM,
publisher=pubACM,
}

@misc{plotly,
  author = {Plotly},
  title = {Plotly Express},
  howpublished = {\url{https://plotly.com/python/plotly-express/}},
  year={2025}
}

@book{python,
  title={Python 3 Reference Manual},
  author={Van Rossum, Guido and Drake, Fred L.},
  year={2009},
  publisher={CreateSpace},
  url={https://docs.python.org/3/}
}

@article{kuksenok2017deliberate,
  title={Deliberate Individual Change Framework for Understanding Programming Practices in Four Oceanography Groups},
  author={Kuksenok, Kateryna and Aragon, Cecilia and Fogarty, James and Lee, Charlotte P and Neff, Gina},
  journal={Computer Supported Cooperative Work},
  volume={26},
  number={4},
  pages={663--691},
  year={2017},
  publisher={Springer},
  doi={https://doi.org/10.1007/s10606-017-9285-x}
}

@article{Perkel2018computational,
author = {Jeffrey Perkel},
year = {2018},
pages = {145--146},
title = {Why Jupyter is Data Scientists' Computational Notebook of Choice},
volume = {563},
journal = {Nature},
doi = {10.1038/d41586-018-07196-1}
}

@article{perkel2021reactive,
    author={Jeffrey Perkel},
    year={2021},
    title={Reactive, Reproducible, Collaborative: Computational Notebooks Evolve},
    journal = jourNat,
    address = addrNat,
    publisher = pubSpringerNat,
    volume = {593},
    pages = {156--157},
    doi={10.1038/d41586-021-01174-w},
}

@misc{Openai, 
    title={GitHub - openai/whisper: Robust Speech Recognition via Large-Scale Weak Supervision},
    url={https://github.com/openai/whisper}, 
    journal={GitHub}, 
    author={OpenAI}, 
    language={en},
    year={2025}
    
}

@inproceedings{midas2020nanoscience,
author = {Nouwens, Midas and Borowski, Marcel and Fog, Bjarke and Klokmose, Clemens Nylandsted},
title = {Between Scripts and Applications: Computational Media for the Frontier of Nanoscience},
  booktitle=procCHI,
  address=addrACM,
publisher=pubACM,
year = {2020},
doi = {10.1145/3313831.3376287},
pages = {1–13},
}

@misc{ryanSemiFormal, 
    author={Ryan Yen},
    title={Semi-Formal Programming, prepared for CHI 2025 Tools for Thought Workshop},
    url={https://ryanyen2.github.io/publications/semi-formal-programming/},
    month=jun,
    year={2025},
    language={en} 
}

@software{Litt_Schoening_Shen_Sonnentag, 
    title={Potluck}, 
    url={https://www.inkandswitch.com/potluck/demo/?openDocument=plant-watering}, 
    journal={Ink & Switch}, 
    author={Litt, Geoffrey and Schoening, Max and Shen, Paul and Sonnentag, Paul}, 
    year={2025} 
}

@misc{malleablesoftware,
 title={Malleable software: Restoring user agency in a world of locked-down apps}, 
 author={Geoffrey Litt and
Josh Horowitz and
Peter van Hardenberg and
Todd Matthews},
 url={https://www.inkandswitch.com/essay/malleable-software/},
  year={2025}
}

@misc{Ruiz_2023, 
    title={make real, the story so far}, 
    url={https://tldraw.substack.com/p/make-real-the-story-so-far}, 
    journal={tldraw}, 
    author={Ruiz, Steve}, 
    year={2023}, 
    month=nov, 
    language={en} 
}

@misc{Victor_2010, 
    title={Ten Brighter Ideas? An explorable explanation}, 
    url={https://worrydream.com/TenBrighterIdeas/}, 
    author={Victor, Bret}, 
    year={2010}, 
    month=mar, 
    language={en} 
}

@article{braun2021one,
  author    = {Virginia Braun and Victoria Clarke},
  title     = {One Size Fits All? What Counts as Quality Practice in (Reflexive) Thematic Analysis?},
  journal   = {Qualitative Research in Psychology},
  volume    = {18},
  number    = {3},
  pages     = {328--352},
  year      = {2021},
  publisher = {Taylor \& Francis},
  doi       = {10.1080/14780887.2020.1769238}
}

@article{braun2006using,
  author    = {Virginia Braun and Victoria Clarke},
  title     = {Using Thematic Analysis in Psychology},
  journal   = {Qualitative Research in Psychology},
  volume    = {3},
  number    = {2},
  pages     = {77--101},
  year      = {2006},
  publisher = {Hodder Arnold},
  address   = {United Kingdom},
  doi       = {10.1191/1478088706qp063oa}
}

@article{mogensen1992towards,
  title={Towards a Provotyping Approach in Systems Development},
  author={Mogensen, Preben},
  journal={Scandinavian Journal of Information Systems},
  volume={4},
  number={1},
  pages={5},
  year={1992},
  url={http://aisel.aisnet.org/sjis/vol4/iss1/5}
}

@inproceedings{hannay2009scientists,
    title={How do Scientists Develop and Use Scientific Software?},
    author={Hannay, Jo Erskine and MacLeod, Carolyn and Singer, Janice and Langtangen, Hans Petter and Pfahl, Dietmar and Wilson, Greg},
    booktitle={Proceedings of the International Workshop on Software Engineering for Computational Science and Engineering},
    pages={1--8},
    year={2009},
    address = addrIEEE,
    publisher = pubIEEE,
    doi={10.1109/SECSE.2009.5069155}

}

@inproceedings{satyanarayan2014declarative,
  title={Declarative Interaction Design for Data Visualization},
  author={Satyanarayan, Arvind and Wongsuphasawat, Kanit and Heer, Jeffrey},
  booktitle=procUIST,
  address=addrACM,
  publisher=pubACM,
  pages={669--678},
  year={2014}
}

@article{bostock2011d3,
  title={D$^3$ Data-Driven Documents},
  author={Bostock, Michael and Ogievetsky, Vadim and Heer, Jeffrey},
  journal={IEEE Transactions on Visualization and Computer Graphics},
  volume={17},
  number={12},
  pages={2301--2309},
  year={2011},
  address=addrIEEE,
  publisher=pubIEEE,  
  doi={10.1109/TVCG.2011.185}
}

@software{react,
    author={{Meta}}, 
    title={React},
    year= {2025}
}

@inproceedings{berger2024scientists,
  title={Scientists and Code: Programming as a Tool},
  author={Berger, Caroline and Lutze, Matthew and Elmqvist, Niklas and Madsen, Magnus and Klokmose, Clemens Nylandsted},
  organization={14th Annual Workshop at the
Intersection of Programming Languages and Human-Computer Interaction},
    year={2024},
    doi={https://doi.org/10.1184/R1/25587726.v1}
}

@inproceedings{ravipreliminary,
  title={A Preliminary Study of the Needs of Climate Data Users},
  author={Ravi, Savitha and Gauravarapu, Sai Krishna Ranjan and Coblenz, Michael},
  organization={14th Annual Workshop at the
Intersection of Programming Languages and Human-Computer Interaction},
    year={2024},
    doi={https://doi.org/10.1184/R1/25586877.v1}
}

@article{kelly2015scientific,
  title={Scientific Software Development Viewed as Knowledge Acquisition: Towards Understanding the Development of Risk-averse Scientific Software},
  author={Kelly, Diane},
  journal={Journal of Systems and Software},
  volume={109},
  pages={50--61},
  year={2015},
  publisher={Elsevier},
  doi={https://doi.org/10.1016/j.jss.2015.07.027}
}

@article{easterbrook2009engineering,
  title={Engineering the software for understanding climate change},
  author={Easterbrook, Steve M and Johns, Timothy C},
  journal={Computing in science \& engineering},
  volume={11},
  number={6},
  pages={65--74},
  year={2009},
  address=addrIEEE,
  publisher=pubIEEE,  
  doi={https://doi.org/10.1109/mcse.2009.193}
}

@inproceedings{widder2019barriers,
  title={Barriers to Reproducible Scientific Programming},
  author={Widder, David Gray and Sunshine, Joshua and Fickas, Stephen},
  booktitle={IEEE Symposium on Visual Languages and Human-Centric Computing},
  pages={217--221},
  year={2019},
  doi={https://doi.org/10.1109/vlhcc.2019.8818907},
  address=addrIEEE,
  publisher=pubIEEE,
}

@software{tableau, 
    author={{Salesforce}}, 
    title={Tableau},
    year=2025
}

@software{Rlang, 
    author={{R Project}}, 
    title={R}, 
    year=2025
}

@article{munzner2009nested,
  title={A Nested Model for Visualization Design and Validation},
  author={Munzner, Tamara},
  journal=jourTVCG,
  volume={15},
  number={6},
  pages={921--928},
  year={2009},
  address=addrIEEE,
  publisher=pubIEEE,
  doi={https://doi.org/10.1109/TVCG.2009.111}
}

@inbook{thomson2014data,
  title={Data Analysis Methods in Physical Oceanography},
  author={Thomson, Richard E and Emery, William J},
  year={2014},
  publisher={Elsevier},
  chapter={2},
  doi={http://dx.doi.org/10.1016/B978-0-12-387782-6.00002-8}
}

@inproceedings{arawjo2020write,
  title={To Write Code: The Cultural Fabrication of Programming Notation and Practice},
  author={Arawjo, Ian},
  booktitle=procCHI,
  pages={1--15},
  year={2020},
  doi={https://doi.org/10.1145/3313831.3376731}
}

@inproceedings{pollock2024designing,
  title={Designing for Semi-Formal Programming with Foundation Models},
  author={Pollock, Josh and Arawjo, Ian and Berger, Caroline and Satyanarayan, Arvind},
  booktitle={14th Annual Workshop at the
Intersection of Programming Languages and Human-Computer Interaction},
  year={2024},
  url={https://vis.csail.mit.edu/pubs/semi-formal-design-space.pdf}
}

@book{disessa_changing_2001,
    title = {Changing Minds: {Computers}, Learning, and Literacy},
    publisher = {MIT Press},
    author = {DiSessa, Andrea A},
    year = {2001},
}

@inproceedings{afzal2019state,
  title={The state of the art in visual analysis approaches for ocean and atmospheric datasets},
  author={Afzal, Shehzad and Hittawe, Mohamad Mazen and Ghani, Sohaib and Jamil, Tahira and Knio, Omar and Hadwiger, Markus and Hoteit, Ibrahim},
  booktitle={Computer graphics forum},
  volume={38},
  number={3},
  pages={881--907},
  year={2019},
  organization={Wiley Online Library}
}

@inproceedings{janicke2020participatory,
  title={Participatory Visualization Design as an Approach to Minimize the Gap between Research and Application.},
  author={J{\"a}nicke, Stefan and Kaur, Pawandeep and Kuzmicki, Pawel and Schmidt, Johanna},
  booktitle={VisGap@ Eurographics/EuroVis},
  pages={35--42},
  year={2020}
}

@article{ulmer2023survey,
  title={A survey on progressive visualization},
  author={Ulmer, Alex and Angelini, Marco and Fekete, Jean-Daniel and Kohlhammer, J{\"o}rn and May, Thorsten},
  journal={IEEE Transactions on Visualization and Computer Graphics},
  volume={30},
  number={9},
  pages={6447--6467},
  year={2023},
  publisher={IEEE}
}

@inproceedings{ahlberg1992dynamic,
  title={Dynamic queries for information exploration: An implementation and evaluation},
  author={Ahlberg, Christopher and Williamson, Christopher and Shneiderman, Ben},
  pages={619--626},
  year={1992},
  booktitle = procCHI,
  publisher = pubACM,
  address   = addrACM,
}

\end{document}